\documentclass[11pt]{article}
\usepackage{geometry}                
\usepackage{graphicx}
\usepackage{amssymb, amsmath}
\usepackage{epstopdf}
\usepackage{natbib}
\usepackage{color}
\usepackage{multirow}
\title{The Linear Lasso: a location model resolution}
\author{D.A.S. Fraser and M. B\'{e}dard}

\begin{document}
\maketitle

\maketitle \noindent\hrulefill
\begin{abstract}
We use location model methodology to guide the least squares analysis of the Lasso problem of variable selection and inference. The nuisance parameter is taken to be an indicator for the selection of explanatory variables and the interest parameter is the response variable itself.  Recent theory eliminates the nuisance parameter by marginalization on the data space and then uses the resulting distribution for inference concerning the interest parameter.  We develop this approach and find: that primary 
inference is essentially one-dimensional rather than $n$-dimensional; that inference focusses on the response variable itself rather than  the least squares estimate (as variables are removed); that first order probabilities are available; that computation is relatively easy;  that  a scalar marginal model is available; and that  ineffective variables can be removed by distributional tilt or shift. 
\end{abstract}
\noindent\hrulefill \\

\textbf{AMS 2000 subject classifications}: Primary ?????; secondary ?????.


\section{Overview}

The Lasso problem concerns data on many variables, one of which is of particular importance, and seeks a small selection of  other variables that give good linear prediction of the important variable. The familiar standardization used is location scale for each variable; we add sign standardization so all the explanatory variables as used are positively correlated with the interest variable; this makes the problem more visualization friendly. We also modify the usually chosen objective function to more closely agree with its intended purpose, although effectively unchanged in the saturated cases. As a consequence the elimination of ineffective variables becomes easier avoiding the usual iteration procedure and making the problem largely dimension free. We use the term Linear Lasso for our procedure to emphasize that the minimization trajectory above the parameter space is a straight line in contrast to the regular Lasso that has multiple segments of lines and curves. 

In \S\ref{section:intro} we review the development and in \S\ref{section:notation} record background and notation. The stochastic background is analyzed in \S\ref{section:LSM} from a geometrical viewpoint, and is linked to location model theory. Section \ref{section:dist} introduces what is viewed as the latent or simulation model and shows that least squares is effectively equivalent to routine Normal analysis, with a very simple example given in the subsequent section. Section \ref{section:y-content} determines how much response distribution is hidden in a selection of explanatory variables and 
records the corresponding selection model. Sections \ref{section:reduced} and \ref{section:algo} show how to construct a reduced set of explanatory variables, while \S\ref{section:crime} and \S\ref{section:grades} illustrate the theory with two real data examples. We conclude with a discussion in \S\ref{section:discussion}.



\section{Introduction} \label{section:intro}

The Lasso (least absolute shrinkage and selection operator) approach is a regression method that simultaneously performs variable selection and parameter estimation. Introduced in the statistical literature by \cite{tibshirani:1996}, its goal is to enhance the accuracy of predictions while retaining the interpretability aspect of the resulting statistical model. The idea behind Lasso is to force the sum of the absolute regression coefficients to be smaller than a predetermined value, which consequently forces some coefficients to be null. It was initially introduced in the context of linear regression and least-squares estimation but its applicability is much wider including, for instance, generalized linear models and proportional hazards models.

The approach essentially consists in a constrained minimization problem over the possible regression coefficients. The criterion to minimize may vary in different contexts, but the constraint on the sum of absolute regression coefficients is generally present (although variations of this constraint may be used in different versions of Lasso). The geometric interpretation of the Lasso is usually illustrated by comparing the shape of its constraint region to that arising from other penalty functions.

In this paper, we fine-tune the objective variable of the standard Lasso and obtain an iteration-free version of the procedure; this resolution of the Lasso is essentially explicit, with performance well exceeding that of the regular Lasso. Specifically, we focus on normal linear models as a way to handle least squares and rely on geometry to propose a simple resolution for the Lasso problem. 
The response variable serves as the focal point of the interpretation, around which explanatory variables gravitate. 
The angles between the response variable and explanatory variables, and among  pairs of explanatory variables, provide the basic input for a geometric analysis guided by location model theory. The response variable is taken as primary and an indicator function is used for selection of explanatory variables. We then examine how well a selection can estimate the response distribution given by prediction variance. We eliminate seemingly underperforming explanatory variables by a tilt or moment generating type modification and avoid negative coefficients under the distributional shift. 

\section{Background and notation} \label{section:notation}

In its original formulation, the Lasso problem considers a scalar variable $y$ of particular interest and $r$ potential explanatory scalar variables $x_1,\ldots,x_r$ typically with $r$ large. 
It then seeks a small sub-selection of  the explanatory variables,  say with subscripts in   $J_s =\{j_1, \ldots , j_s\}$, that alone provides acceptable or good prediction for   the $y$ value with new data. To perform this task we have available $n$ observations on the $1+r$ variables, giving  
 full data   as an  $n \times (1+r)$ array $({\bf y}, {\bf x}_1, \ldots , {\bf x}_r)$ or  as $1+r$ vectors of length $n$. The location, scaling, sign, and  more of the variables are typically conventional so we can widely apply standardizations. 
Accordingly, we hereafter assume that each column vector has been location-scale standardized so the average of the coordinates is zero and the standard deviation  is one.  And then to keep notation simple we, typically, use the same notation   $({\bf y},  {\bf x}_1, \ldots, {\bf x}_r)$ for the modified data. 

An important  objective of  Lasso is  to estimate or predict the value of the variable  $y$ that corresponds  to  the observed  values of 
the selected explanatory variables. 
The usual Lasso procedure for given data  is to minimize, over choice of regression coefficients $\beta$, the expression
$$
\sum_i (y_i-  {\bf X}_i \beta)^2 /2n + \gamma \Sigma_j |\beta_j| \ ,
$$
where ${\bf X}_i$ is the $i$-th row of ${\bf X}=({\bf x}_1, \ldots , {\bf x}_r)$. The first term is a rescaled sum-of-squares of departures from the linear model and the second term is a Lagrangian restraint or penalty to force fewer selected predictors, with $\gamma$ for tuning; see 
\cite{HastieTW:2015}. 
The few non-null regression coefficients $\beta$ retained by the Lasso can then be combined with new observations from the selected explanatory variables to predict the associated response $y$. 

Following the previous standardization step, each data vector  
becomes a vector of length $n^{1/2}$ and  correlations are  obtained  by dividing inner products by $n$. We view the correlations as the intrinsic data for the problem; let  $c=(c_j)$  be the correlations between $\bf{y}$ and the ${\bf x}_j$ vectors, and let $C=(c_{jk})$ be the correlations among the ${\bf x}_j$  vectors.
 We can then assemble these  as a  full matrix of correlations  among all the  vectors 
\begin{eqnarray*}
\tilde{C} &=& \left(
\begin{array}{cc}
1 & c^t \\
c & C 
\end{array}
\right) \ .
\end{eqnarray*}
The use of the letters $c$ and $C$   is to remind  that the elements are just cosines of 
 angles among    unit  data vectors, each conveniently obtained from a corresponding  inner product. As with Lasso, the correlations are treated as first order constant.

\begin{figure}[hbt]
	\centering
	\includegraphics[scale=0.67, angle=0, trim={20 110 0 100}, clip] {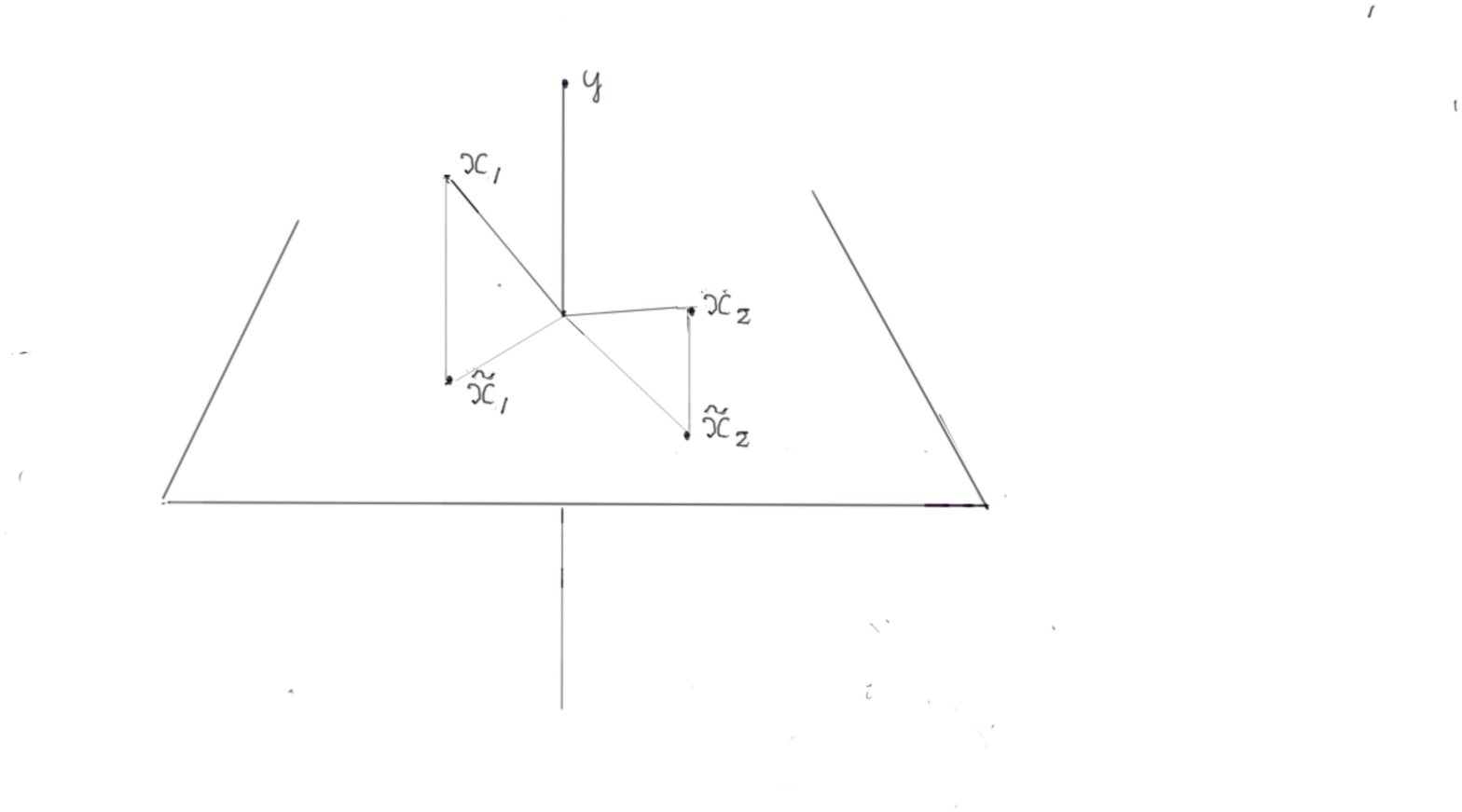} 
    	\caption{The response vector ${\bf y}$ and two predictor vectors ${\bf x}_1$ and ${\bf x}_2$, along with their projections $c_i$ to $       {\mathcal L}	\bf y$ and residuals  $\tilde{\bf x}_i$ on ${\mathcal L}^{\perp}\bf y$}.
	\label{figure:vectors}
\end{figure}

As explanatory variables ${\bf x}_j$ can be positively or negatively correlated with the response ${\bf y}$ we choose a further standardization, one for visual convenience: any explanatory variable that has a negative correlation 
with the response vector has its sign reversed. Then, all explanatory vectors are positively correlated with the response vector.
This modification is notational and for visualization only, and does not affect the substance; indeed it
is in some agreement with usual regression analysis.
It is also natural to have the zero point of each vector placed directly on the origin of some underlying vector space,
and even
to view ${\bf y}$ as
pointing upward. Then  all the ${\bf x}_j$ vectors   would   be  directed  into the upper half space ${\mathcal L}^{+}\bf y$, above the plane $\mathcal{L}^{\perp}\bf y$
perpendicular to $\bf y$; see Figure \ref{figure:vectors}.

\section{Latent stochastic model} \label{section:LSM}
The reference to correlated data indicates a common stochastic background for the $1+ r$ variables.  As part of this each variable  can be  viewed to   first-order as a linear function of a  latent  standard  Normal distribution on an  underlying vector space.  The values of such a linear function can be recorded on the line perpendicular to the contours of the linear function, that is on the lines  formed by the vectors $\mathbf{y},\mathbf{x}_1,\cdots, \mathbf{x}_r$  and designated   ${\mathcal L}{\bf y}, {\mathcal L} {\bf x}_1, \cdots, {\mathcal L}{\bf x}_r$. 
Conveniently, the use of least squares has algebraic equivalence to  symmetric Normal location-model analysis where  the related distribution theory  provides important guidance.

As part of this we 
introduce  a further coordinate standardization that makes model form and  objectives more transparent, and then
 discuss the latent location model itself. As indicated above we have  that the
  $y$ values from a  linear function can 
  be recorded  on the  line 
${\mathcal L}{\bf y}$
in the  latent variable space; prediction is thus one-dimensional rather than $n$-dimensional, and the analysis thus involves  scalar fitting rather than $n$-dimensional regression fitting.  This then leads to the Linear Lasso procedure that allows  selection of variables by  minimum-number or  maximum-variance viewpoints, where the elimination of unproductive  $x$ variables is by  tilting  or equivalently by shifting.

For a first-order model that corresponds to least squares, we use an   $n$-dimensional latent variable space having  a rotationally symmetric standard Normal distribution which for convenience we  center at the origin.  
 An  observable variable is then  a linear function on   that space, a sort of ``tap" on the  latent stochastics. 
Each linear function has  its own linear contours on the latent variable space; and then perpendicular to these contours is a line through the origin  that can index  the contours of the function. Each such line  records   values for the corresponding   variable and thus presents  a column  of the given data array.
As the zero points of the vectors are conveniently      placed on  the zero point of the latent vector space $\mathbb{R}^n$, the model gives data on a rotationally-symmetric Normal latent model   using $c$ and $C$ that provide a type of skewed coordinates. The data can then be viewed as giving $n$ values on each line, corresponding  to say successive time points or other.

In the presence of the  latent stochastic model we are able to describe variables of  interest in terms of their dependence on the  
latent variables, that is present them as   ``taps" on the latent stochastics or equivalently as
 functions on the latent variable space. This alternative modelling format  offers  advantages including making explicit the 
  continuity that is  present among variables. Such functions on the latent space can be called ``data
 generating" for their availability for  simulations or ``structural" for their explicit presentation of the dependences.
 For the full set of variables we use the data generating format and a choice of expressive but nonstandard notation; 
$\{y{\bf y}, x_1{\bf x}_1,\cdots, x_r{\bf x}_r \}$ where bold face letters ${\bf y}, {\bf x}_1,\cdots, {\bf x}_r $ are used  for the fixed unit vectors that record  the  directions of the stochastic ``taps",
and  the coefficients $y, x_1, \cdots, x_r$  are  then each standard Normal but collectively have
correlations recorded as $\tilde C$. 
Then for the  modelling  to structure  least squares we use the  lower case  variables which are  jointly  multivariate Normal (0; $\tilde C$), that is
\begin{eqnarray}
 y, {x}_{1},~ \ldots ~, {x}_{r} &\sim& \mathcal{MN} \left(  \left( 
\begin{array}{c}
0 \\
\vdots \\

0 
\end{array}
\right) ; \left(
\begin{array}{ll}
1 & c^t \\
c & C 
\end{array}
\right) \right) \ . \label{joint_y1r}
\end{eqnarray}

\section{Inference from a particular subset of exploratory variables } \label{section:dist}

Now consider the case with some specific subset of $s$ exploratory variables with indices in   $J{_s}=\{ j_1,\cdots, j_s\}$, where for example $j_1$ is the original subscript of the first coordinate selected and so on. 
Then using the multivariate Normal in (1) to manage least squares for the variables labelled by $J_s$, we obtain the distribution
\begin{eqnarray}
 y, {x}_{j_1},~ \ldots ~, {x}_{j_s} &\sim& \mathcal{MN} \left(  \left( 
\begin{array}{c}
0 \\
\vdots \\

0 
\end{array}
\right) ; \left(
\begin{array}{ll}
1 & c^t_{s}\\
c_{s}& C_s
\end{array}
\right) \right) \ ,
\end{eqnarray}
where $c_s$ and $C_s$  designate the correlations restricted to the subset  $J_s$. 

For convenience now, we assume that  the full correlation matrix  $\tilde C$ is nonsingular, and   return  later for greater generality. We then  use the formulas of conditional probability to obtain the  distribution for the  inherent   $y$ content in the exploratory variables  having subscripts in the set $J_s$. For a single $x$ variable the familiar conditioning formula is 
  $y|x \sim \mathcal{N}(\sigma_{y,x} \sigma_{x,x}^{-1} x~ ;~\sigma_{y,y}-\sigma_{y,x} \sigma_{x,x}^{-1} \sigma_{x,y})$; then applying here the corresponding vector version  gives
\begin{eqnarray}
y | x_{j_1}, \ldots, x_{j_s} \sim \mathcal{N} \left(  c^t_{s}C_{s}^{-1}  \left( 
\begin{array}{c} x_{j_1} \\
\vdots \\
x_{j_s}\end{array}
\right) ;1- c^t_{s} C_{s}^{-1} c_{s} \right) \ . \label{prediction}
\end{eqnarray}
This forecasts the value $c^t_{s}C_{s}^{-1}  (x_{j_1} ,\cdots,
x_{j_s})^t$ for the   $y$ content  and this  in turn has standard deviation $\{c^t_{s} C_{s}^{-1} c_{s}\}^{1/2}$. This standard deviation gives the $y$ effect inherent in the exploratory variables  with subscripts in the set $J_s$.

\section{A very simple example} \label{section:simple_ex}
Consider a very simple example as indicated by Figure 1, with $n=3$ and $r=2$. In that context, the number of possible selected variables is  $s=1$ or $s=2$. Suppose the data array is
\begin{eqnarray*}
({\mathbf y}~ \mathbf{X}) = (\mathbf y~ \mathbf{x}_1~ \mathbf{x}_2) = \left( 
\begin{array}{lll}
1.000\ 000 & 0.900\ 000 & 0.600\ 000 \\
0.0 & 0.435\ 890 & 0.400\ 000  \\
0.0 & 0.0 & 0.692\ 820
\end{array}
\right) \ ;
\end{eqnarray*}
the corresponding correlations are 
\begin{eqnarray*}
\tilde{C} =  \left(
\begin{array}{cc}
1 & c^t \\
c & C 
\end{array}
\right)
&=& \left( 
\begin{array}{lll}
1.000\ 000 & 0.900~000 & 0.600~000 \\
0.900~000 & 1.000\ 000 & 0.714\ 356  \\
0.600~000 & 0.714\ 356 & 1.000\ 000
\end{array}
\right) \ .
\end{eqnarray*}
 
The latent stochastic model can then be expressed as $(y \mathbf{y}, x_1 \mathbf{x}_1,~ x_2 \mathbf{x}_2)$, where $(y, x_1, x_2)$ is a multivariate Normal $(0, \tilde{C})$ given the directions $(\mathbf{y},  \mathbf{x}_1,   \mathbf{x}_2)$. 
In Figure \ref{figure:vectors}, the latent model is a unit standard Normal centered at the 
 origin. The form of the model above ${\mathcal L}^{\perp}\bf y$ has a near reflection through the origin, giving a near duplicate below ${\mathcal L}^{\perp}\bf y$ (model has no cubic terms). 

For the example, we only have 3 choices for a selection of explanatory variables, namely $\{ x_1\}$, $\{ x_2\}$, $\{ x_1, x_2\}$. 
Then, for the three possible cases for the variable selection process, general expressions for the $y$-content and the percentage of explained response variability are provided in Table \ref{table:simple_1}. The last line in the table uses results for the full set of explanatory variables.

\begin{table}
\begin{subequations}
\begin{alignat}{3}
&\textnormal{Source } && \textnormal{$y$ content} && \textnormal{SD of content = fraction of $y$ variability}
 \nonumber \\*[5pt] \hline \rule[5pt]{0pt}{10pt}
&\{x_1\} && c_1~x_1 && c_1 \\*[5pt]
&\{x_2\} && c_2~  x_2 && c_2 \\*[5pt]
&\{x_1, x_2\} &\hspace{.2in}
& (c_1~ c_2) \left(
\begin{array}{ll}
1 & c_{12} \\
c_{21} & 1 
\end{array}
\right)^{-1}\left( 
\begin{array}{l}
x_1 \\
x_2
\end{array}
\right)
&\hspace{.2in}&  \left\{ (c_1~ c_2) 
\left(
\begin{array}{ll}
1 & c_{12} \\
c_{21} & 1 
\end{array}
\right)^{-1}
\left( 
\begin{array}{l}
c_1 \\
c_2
\end{array}
\right) \right\}^{1/2}
\end{alignat}
\end{subequations}
\caption{General expressions for individual predictions  (second column) and their SD, representing the fraction of  
$y$-variability (third column), for each possible subset of selected predictors} \label{table:simple_1}
\end{table}

With the data for the simple example summarized in $\tilde{C}$, the general expressions in Table \ref{table:simple_1} lead to the models and values in Table \ref{table:simple_2}. We then see that $\mathbf{x}_1$ has the largest projection on the $\mathbf{y}$ axis, and thus the variable $x_1$ is associated with the largest SD of $y$-content; including $x_2$ then adds very little.  
 This is unsurprising considering that ${\bf x}_2$ is more correlated with ${\bf x}_1$ than it is with ${\bf y}$.
%
%


\begin{table}
\begin{eqnarray*}
\begin{array}{lccc}
\textnormal{Source } &\begin{array}{c} \mathbf{y} \textnormal{ axis} \\ \textnormal{ projection} \end{array} & \textnormal{$y$ content} &\begin{array}{c} \textnormal{ SD } \\ \textnormal{of content} \end{array}   \\*[5pt] \hline \rule[5pt]{0pt}{10pt}
\textnormal{From } \{x_1\} & 0.9 &0.9~x_1 & 0.9 
 \\*[5pt]
\textnormal{From } \{x_2\} & 0.6& 0.6~x_2  &  0.6 
\\*[5pt]
\textnormal{From } \{x_1, x_2\} & 0.902 & 0.963~x_1 - 0.088~x_2  
 &   0.902 
\end{array} 
\end{eqnarray*}
\caption{$y$-content (second column) and SD = fraction of explained 
$y$-variability (third column) for each possible subset of selected predictors} \label{table:simple_2}
\end{table}

Figure 2 provides an illustration of the quantity of $y$-information that is available from each explanatory variable separately. Figure \ref{figure:example2} depicts the quantity of $y$-information that is available from the best selection of size 1 and 2, respectively. In both cases, we included the full $y$ density for comparison.  
Although the example is ultra simple it does illustrate that for any subset of explanatory variables there is a formula for the  information  concerning $y$ that is available from such a  selection. But here the number of possible non-trivial subsets is $r! / \{s!(r-s)!\}=3!/\{2!1!\}=3$, while with larger data arrays the number grows exponentially.

\section{How much  $y$ distribution is hidden in   selected $x$ variables} \label{section:y-content}

Now suppose we  address the larger question of how much  $y$-distribution  information  is accessible from a  specified $J{_s}=\{ j_1,\cdots, j_s\}$ selection of explanatory variables. The distribution of such variables can be presented in data-generating form as $(x_{j_1}{\bf x}_{j_1},\cdots, x_{j_s}{\bf x}_{j_s})$, where the  coefficients ${x}_{j_1}, \ldots, {x}_{j_s}$ are centered multivariate Normal
\begin{eqnarray}
{x}_{j_1}, \ldots, {x}_{j_s}\sim& \mathcal{MN} \left(  \left( 
\begin{array}{c}
0 \\
\vdots \\
0 
\end{array}
\right); C_s = \left(
\begin{array}{cccc}
1 & c_{j_1j_2} & \ldots & c_{j_1j_s}\\
c_{j_2j_1} & 1 & & c_{j_2j_s}\\
\vdots &  & \ddots & \vdots  \\ 
c_{j_sj_1} & \ldots & c_{j_{s-1}j_s }& 1 \\
\end{array}
 \right) \right) \ . \label{eq:marginal-x}
\end{eqnarray}

\begin{figure}[hbt]
	\centering
	\includegraphics[scale=0.67, angle=0, trim={50 72 0 30}, clip] {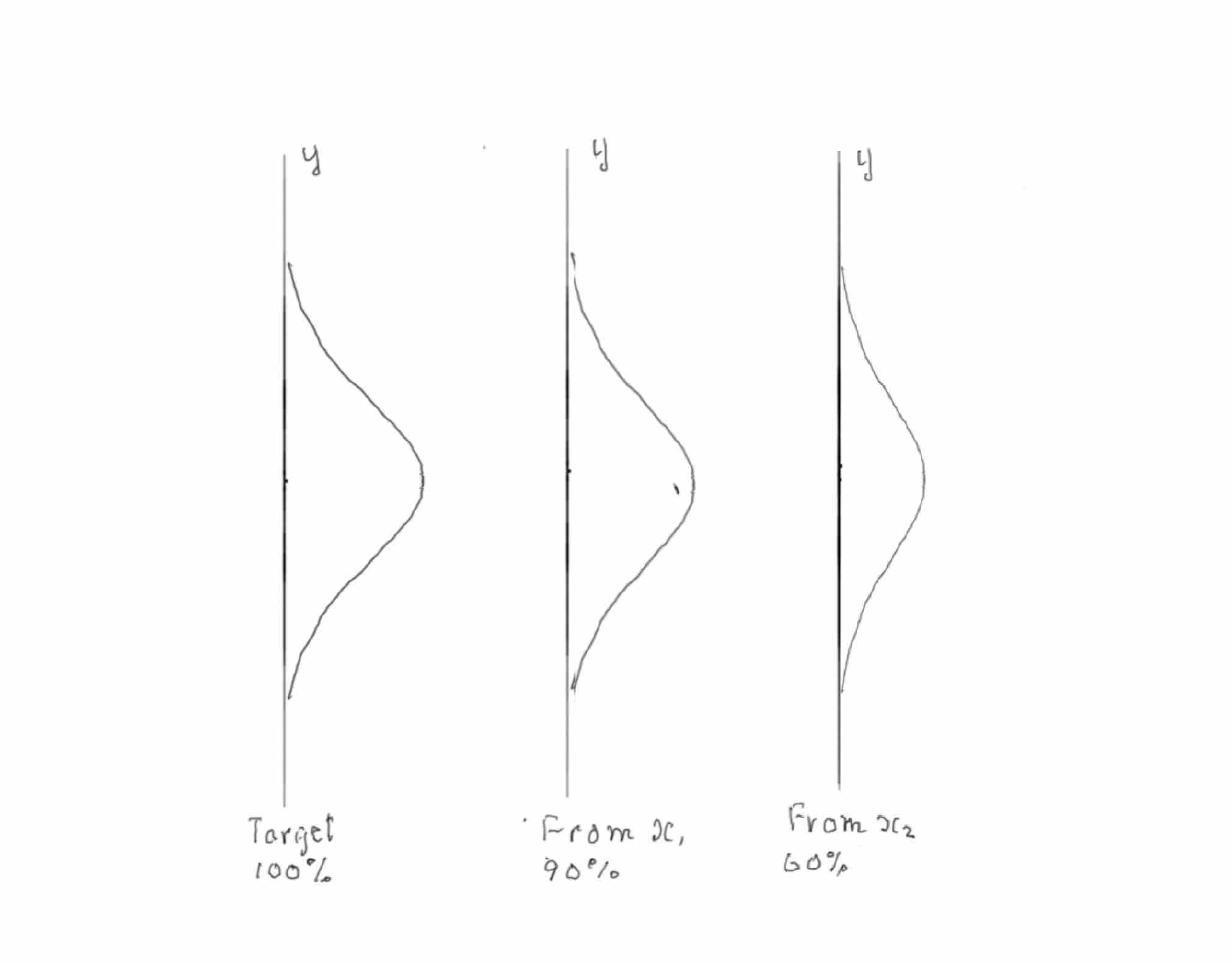}
    	\caption{For the simple example in \S\ref{section:simple_ex} we record the amount  $0.36$ of $y$ density inherent  
	in  $x_2$ (right),  $0.81$ in $x_1$ (middle); and finally $1$ in the target variable $y$ (left); percentages are illustrated by fractioning the height of the standard  normal.} 	
	\label{figure:example} 
\end{figure}
\begin{figure}[hbt]
	\centering
	\includegraphics[scale=0.67, angle=0, trim={30 109 0 70}, clip] {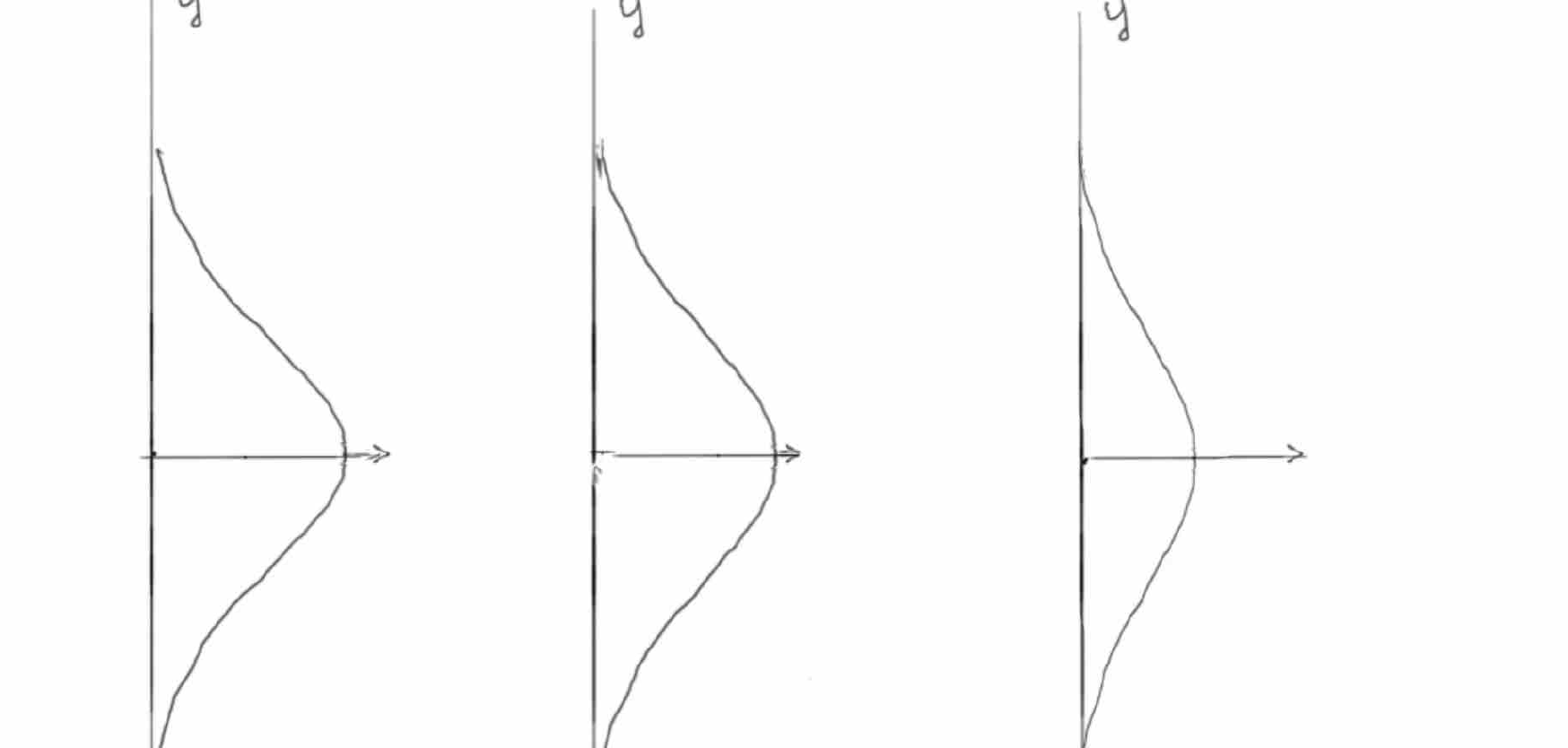}
    	\caption{For the simple example in \S\ref{section:simple_ex} we record the inherent $y$ content $0.81$ from $x_1$ alone (right), $0.814$ from $x_1$ and $x_2$ (middle), and finally $1$ for the target variable $y$ (left). 
	} 
 \label{figure:example2}
\end{figure}

The elements can then  be combined and the result is available in the last paragraph of \S\ref{section:dist}; we have  a centered  Normal  with SD  $\sigma = \{ c^t_{s}C_{s}^{-1} c_{s} \}^{1/2}$.
Thus the $y$-content of the $J_s=\{j_1,\cdots, j_s\}$ selection of explanatory variables is just the   
 fraction 
$\sigma= \{ c^t_{s}C_{s}^{-1} c_{s} \}^{1/2}$
of the probability in the  target   standard Normal distribution of the variable $y$ itself; see Figure \ref{figure:example}.

We are interested in the   $y$-distribution content available from a specific choice $J_s$  of explanatory variables. Towards this
let $\delta=(\delta_1, \cdots, \delta _r)$ be an indicator variable for the presence $(1)$  or absence $(0)$ of each of the $r$ available   explanatory variables. Then from the preceding paragraph the   $y$-distribution content is given as a fraction $\sigma(\delta)$ of a central Normal distribution,  with $\sigma(\delta)= \{ c^t_{\delta}C_{\delta}^{-1} c_{\delta} \}^{1/2}$.  We then have that
the statistical density description of this    information   from  the specified explanatory variables is
\begin{eqnarray}
\sigma(\delta)~~{\rm
 of}~~\phi (y)~~ \equiv~~ \sigma(\delta)~~{\rm of}~~\frac{1}{(2\pi)^{1/2}}\exp\{-y^2/2\} \ .
\end{eqnarray}
This presents a relative density recorded as a fraction of a standard Normal distribution; it can also be recorded more formally in
statistical model format:
\begin{eqnarray}
f\{y;\delta\}= \sigma(\delta)\phi (y) \ .
\end{eqnarray}
The density arising from any component or group of components 
labeled say $\delta$ will typically not integrate to 1, as it is recording just the fraction of a hidden $y$ distribution that is accessible from the $x$ variables.

\section{Substantial reduction in the set $J_s$ of explanatory variables} \label{section:reduced}

Our primary objective  is to find   a small selection  $J_s$ of explanatory variables     that collectively 
give
 large $y$ variance and thus high $y$-distribution content.  
This
 involves
 a trade-off between the   variance  
$\sigma^2(\delta)=c^t_{\delta}C_{\delta}^{-1} c_{\delta}$ and  the cardinality $\Sigma\delta_i$ 
of the 
selection.  A direct search  for this has exponential order of computation and could  be viewed as not feasible with large data.

The variance formula for a single $x_i$ is straightforward;  it  gives     the corresponding squared correlation $c_i^2$, and is  thus  immediately available.
But   if  we  seek an additional  explanatory variable the variance 
is typically not the sum $c_1^2 + c_2^2$ of the individual variances, 
but   includes  weights from  the inverse of the  correlation matrix $C$. The variance for the present sum  is $c_1^2 c^{11} + 2 c_1c_2c^{12}+c_2^2c^{22}$, where the $c$'s with two raised  indices are  elements of $C^{-1}$;  the maximum for this search seems  not so easily available.

For this we briefly discuss a method for removing a large batch of underperforming explanatory variables
 and then in the next section fine-tune this approach to obtain  a one-by-one procedure. Our marginal model is a centered Normal $y$ with variance $\sigma^2(\delta)=c^t_{\delta}C_{\delta}^{-1} c_{\delta}$; this model depends on the selection $\delta$ entirely through the scaling  or spread $\sigma$ of its distribution. We examine this marginal distribution on the positive axis and shift the distribution  to the left by an exponential tilt $\exp\{\gamma y\}$, but in doing this 
retain only positive regression coefficients; this  eliminates underperforming $x_i$ 
and provides a distributional analog of the penalty function approach in  \cite{tibshirani:1996}. The $x_i$ eliminated by this process are those with small $c_i$ values. This substantial reduction is easy, entirely based on small correlations, no iteration needed. It is available here because of our direct search for $y$ content rather than  the focussed use of fitted regression.

\section{One-by-one} \label{section:algo}

When few variables are left, we might account for the correlations  
$C$ by eliminating the variable $x_i$ that leads to the smallest decrease in the variance term $\sigma^2(\delta)=c^t_{\delta}C_{\delta}^{-1} c_{\delta}$. Suppose that $m$ variables have been eliminated in the initial step and that the original subscripts of these variables are listed in $\mathcal{M}$. We thus have $r-m$ variables left for the second step; iteratively, we proceed as follows
\begin{enumerate}
\item Initialize $\delta^{(0)}$ such that $\delta_i = 0$ for $i\in \mathcal{M}$ and $\delta_i = 1$ for $i\notin \mathcal{M}$, $i=1, \ldots, r$. 
The cardinality $\Sigma \delta_i$ of the selection $\delta^{(0)}$ is then $r-m$.
\item Let $\delta^{(1)} = \delta^{(0)}$; then set 
$\delta_{k_1} = 0$, 
where $k_1$ minimizes 
\[
\sigma^2(\delta^{(0)}) - \sigma^2(\delta^{(1)}) = c^t_{\delta^{(0)}} C_{\delta^{(0)}}^{-1} c_{\delta^{(0)}} - c^t_{\delta^{(1)}} C_{\delta^{(1)}}^{-1} c_{\delta^{(1)}} \ ;
\]
the cardinality is thus $r-m-1$.
\item Let $\delta^{(2)} = \delta^{(1)}$; then set 
$\delta_{k_2} = 0$, where $k_2$ minimizes 
\[
\sigma^2(\delta^{(1)}) - \sigma^2(\delta^{(2)}) = c^t_{\delta^{(1)}} C_{\delta^{(1)}}^{-1} c_{\delta^{(1)}} - c^t_{\delta^{(2)}} C_{\delta^{(2)}}^{-1} c_{\delta^{(2)}} \ ;
\]
the cardinality is thus $r-m-2$.
\item We repeat these steps until all variables have been removed; the resulting ordering provides a progressive selection of variables for various cardinalities $\Sigma \delta_i$.  
\end{enumerate}

Combining the substantial reduction of \S\ref{section:reduced} with the present one-by-one procedure gives the process for the Linear Lasso; this new approach works directly with the target $y$ change, after the data have been sign standardized. Neither the  Linear nor the regular   Lasso  can  be expected to fully achieve its desired  optimization, but the Linear uses $y$-change directly as the desired target and in addition has the substantial property of avoiding  iterative steps. The tuning parameter for the Linear Lasso is the $\gamma$ for the exponential tilt or shift.

\section{Example: Crime data} \label{section:crime}

To illustrate the use of the Linear Lasso, we study the small example on page 10 of \cite{HastieTW:2015}. The data, originally taken from \cite{thomas:1990}, reports the crime rate per million of residents in $n=50$ U.S.~cities. There are $r=5$ explanatory variables: annual police funding (dollars/resident), people age $\geq 25$ with four years of high-school (\%),  people age 16 to 19 not in high school, nor high-school graduates (\%), people age 18 to 24 in college (\%), and people age $\geq 25$ with $\geq 4$ years of college (\%).  The resulting data array is $50 \times 6$; the crime rate in the first column of the array is the outcome vector 
and columns 2 to 6 are the five potential explanatory variables. 

A first step consists in standardizing the data so that each column has an average of 0 and a standard deviation of 1. The standardized data array is $(\mathbf{y} \ \mathbf{x}_1 \ \ldots \ \mathbf{x}_5)$. Inner products $c_i = \mathbf{y} \cdot \mathbf{x}_i / n$ ($i=1, \ldots, 5$) represent the correlations between the outcome $\mathbf{y}$ and the explanatory variables ($\mathbf{x}$'s). The variables $\mathbf{x}_2$, $\mathbf{x}_4$, $\mathbf{x}_5$ represent different measures of the population's education level, and are all negatively correlated with the crime rate. If we picture the outcome $\mathbf{y}$ as a vector pointing upwards, this implies that vectors $\mathbf{x}_2, \mathbf{x}_4, \mathbf{x}_5$ lie in the lower half space. We invert the sign of these three explanatory variables and use $\mathbf{x}^*_i = -\mathbf{x}_i$ (for $i=2,4,5$) and $\mathbf{x}^*_i = \mathbf{x}_i$ (for $i=1,3$), to instead work with vectors that are in the upper half space $\mathcal{L}^+ \mathbf{y}$. The resulting vector of correlations $c$ is then formed of the terms $|c_i| = \mathbf{y} \cdot \mathbf{x}^*_i / n$ ($i=1, \ldots, 5$). Similarly, inner products $c_{jk} = \mathbf{x}^*_j \cdot \mathbf{x}^*_k / n$ ($j,k=1, \ldots, 5$) form the $5 \times 5$ matrix $C$ of correlations between the $\mathbf{x}^*$ vectors.

From \eqref{joint_y1r}, the latent model $(y, x^*_1, \ldots, x^*_5)$ is jointly distributed according to a $\mathcal{MN}(\mathbf{0}, \tilde{C})$. Using \eqref{prediction}, the predictive distribution for the full model is  
\begin{eqnarray*}
y | x^*_{1}, \ldots, x^*_{5} 
&\sim& \mathcal{N} \left(  0.516 x^*_1 +0.206x^*_2 +0.112 x^*_3  - 0.019 x^*_4 -  0.097 x^*_5, ; 0.666 \right) \ .
\end{eqnarray*}
Of interest now is to find a small subset of explanatory variables that features high $y$-distribution content. A predictive distribution with a small variance (variance about prediction) is hoped for. 
We thus look for a selection $J_s$ of explanatory variables that is associated to a high variance (variance of prediction); the associated standard deviation then 
represents the fraction of $y$-variability that is explained by the model (marginally). 

We proceed as expounded in Section \ref{section:algo}, starting with the full model and iteratively removing variables. In this example, the vector of correlations between the outcome $\mathbf{y}$ and the explanatory variables $\mathbf{x}_i ^*$ is $c^t = (0.533, 0.135, 0.323, 0.175, 0.026)$. Let us suppose that the first $m$ variables are eliminated from the model on the basis of having small $c_i$; the remaining $r-m$ variables then have the highest correlations. This elimination procedure can be viewed as a one-sided version of the two-sided penalty function used in \cite{tibshirani:1996}. And as such it uses a moment generating type penalty with the ``other side", managed by having only non-negative regression coefficients; in turn the moment generating function penalty with the standard normal latent distribution just provided a shift of that distribution. In each step this eliminates the smallest contributor to estimable $y$-distribution. 

The resulting sequence of models is detailed in Table \ref{figure:LL} for different choices of $m$, along with the corresponding percentage of $y$-content for each model. For comparison, Table \ref{table:Lasso} provides the models obtained using Lasso with 
a continuum of $\gamma$-values. Lasso and Linear Lasso do not propose the same sequences of models; in fact, Linear Lasso with $m=3$ and $m=5$ are the only instances with identical sequences. In all cases, the first variable to be eliminated is always either the fifth or fourth one.  

\begin{table}
\begin{center}
\begin{tabular}{lc||ccccc}
 \multicolumn{7}{c}{{\bf Linear Lasso}} \\*[3pt] \hline \rule{0pt}{12pt} 
& $s$ & 5 & 4 & 3 & 2 & 1  \\*[1pt] \hline \rule{0pt}{15pt} 
\multirow{4}{*}{$m=0$}& $J_s$ & \{1,2,3,4,5\} & \{1,2,3,5\} & \{1,2,5\} & \{1,2\} & \{1\}  \\*[5pt]
&  {\% $y$-cont.} & 57.758 & 57.749 & 57.277 & 56.984 & 53.320  \\*[5pt]
& cv-mse & {0.8524} & {0.8476} & {0.8434} & {0.7864}  &   \textbf{0.7784} \\*[5pt] 
&  sd & {0.0397} & {0.0393} & {0.0372} & {0.0273}  &   {0.0452} \\*[3pt] \hline \rule{0pt}{15pt}
\multirow{4}{*}{$m=1$}& $J_s$ & \{1,2,3,4,5\} & \{1,2,3,4\} & \{1,2,3\} & \{1,2\} & \{1\}  \\*[5pt]
&  {\% $y$-cont.} & 57.758 & 57.548 & 57.231 & 56.984 & 53.320  \\*[5pt]
& cv-mse & {0.8556} &  {0.8305} &  {0.8398} & {0.7919} & \textbf{0.7776}  \\*[5pt]
&  sd & {0.0381} &  {0.0331} &  {0.0342} &  {0.0281} &  {0.0420} \\*[3pt] \hline \rule{0pt}{15pt}
\multirow{4}{*}{$m=3$}& $J_s$ & \{1,2,3,4,5\} & \{1,2,3,4\} & \{1,3,4\} & \{1,3\} & \{1\}  \\*[5pt]
&  {\% $y$-cont.} & 57.758 & 57.548 & 56.457 & 55.802 & 53.320  \\*[5pt]
& cv-mse & {0.8582} &  {0.8288} &  {0.7883} &  \underline{\textbf{0.7679}} & {0.7757}   \\*[5pt]
&  sd & {0.0498} &  {0.0380} &  {0.0362} &  {0.0424} &  {0.0389} \\*[3pt] \hline \rule{0pt}{15pt}
\multirow{4}{*}{$m=5$}& $J_s$ & \{1,2,3,4,5\} & \{1,2,3,4\} & \{1,3,4\} & \{1,3\} & \{1\} \\*[5pt]
&  {\% $y$-cont.} & 57.758 & 57.548 & 56.457 & 55.802 & 53.320  \\*[5pt]
& cv-mse &  {0.8554} & {0.8312} & {0.7870} & \textbf{0.7756} & {0.7837}  \\*[5pt] 
& sd & {0.0420} & {0.0373} & {0.0291} & {0.0571} & {0.0613}  \\*[3pt] \hline 
\end{tabular}
\caption{Selection of subsets $J_s$ obtained with different $m$ values in the Linear Lasso, along with their  corresponding fraction of $y$-content ($\{c_{s}^t C_{s}^{-1} c_{s}\}^{1/2}$, in \%). Mean-squared prediction errors and their standard deviations, 
obtained with 50 repetitions of a 10-fold cross-validation, are also provided. 
} \label{figure:LL}
\end{center}
\end{table}

\begin{table}
\begin{center}
\begin{tabular}{r|ccccccccc} 
\multicolumn{10}{c}{{\bf Lasso}} \\*[3pt] \hline \rule{0pt}{12pt} 
$\gamma$ & \footnotesize{0.00} & \footnotesize{0.03} & \footnotesize{0.06} & \footnotesize{0.10} & \footnotesize{0.14} & \footnotesize{0.18} & \footnotesize{0.22} &  \footnotesize{0.25} & \footnotesize{0.30} \\*[3pt]
$s$ & \footnotesize{5} & \footnotesize{4} & \footnotesize{3} & \footnotesize{3} &  \footnotesize{3} & \footnotesize{2} & \footnotesize{2} & \footnotesize{1} & \footnotesize{1} \\*[3pt] \hline \rule{0pt}{15pt}
$J_s$ & \footnotesize{\{1,2,3,4,5\}} & \footnotesize{\{1,2,3,5\}} & \footnotesize{\{1,2,3\}} & \footnotesize{\{1,2,3\}} &\footnotesize{\{1,2,3\}} & \footnotesize{\{1,3\}} & \footnotesize{\{1,3\}} & \footnotesize{\{1\}} & \footnotesize{\{1\}} \\*[5pt] 
\% $y$-cont. &  \footnotesize{57.758} & \footnotesize{57.749} & \footnotesize{57.231} & \footnotesize{57.231} & \footnotesize{57.231} & \footnotesize{55.802} & \footnotesize{55.802} & \footnotesize{53.320}& \footnotesize{53.320} \\*[5pt]  
cv-mse  & \footnotesize{0.8673} & \footnotesize{0.8213} & \footnotesize{0.8037} & \footnotesize{0.8158}  &   \footnotesize{0.8377} & \footnotesize{0.8591} & \footnotesize{0.8935} & \footnotesize{0.9252} & \footnotesize{0.9492}  \\*[5pt] 
sd & \footnotesize{0.04488} & \footnotesize{0.0401} & \footnotesize{0.0362} & \footnotesize{0.0383}  &   \footnotesize{0.0440}&\footnotesize{0.0445}&\footnotesize{0.0439}&\footnotesize{0.0265}&\footnotesize{0.0254} \\*[5pt]
\end{tabular} 
\caption{Mean-squared prediction errors and their standard deviations, 
obtained with 50 repetitions of a 10-fold cross-validation. 
} \label{table:Lasso}
\end{center}
\end{table}

To find out the optimal number of explanatory variables in terms of prediction, we use a
repeated 10-fold cross-validation approach.  The 50 observations are randomly divided into 10 groups of size 5. One of these groups is taken as a test set, while the nine remaining groups form the training set. The Linear Lasso is applied to a training set to obtain a sequence of nested models, as well as estimates associated to these models. Each of the five fitted models is then used to predict responses in the test set; for each model, we record the mean-squared prediction errors. These steps are repeated 10 times, each time taking a different group as the test set.  The mean-squared prediction errors are averaged separately for each of the five models. This process is then repeated 50 times, each time with a new random partitioning of the observations into 10 groups. For each of the 5 models, the output is thus a 50-dimensional vector of mean-squared prediction errors. Table \ref{figure:LL} reports the mean and standard error associated to each model. 
The same steps are then repeated for the standard Lasso, for several values of the tuning parameter $\gamma$; results are reported in Table \ref{table:Lasso}.

According to Table \ref{figure:LL}, the Linear Lasso favors the model with two explanatory variables ($x_1^*, x_3^*$) as this is the selection that minimises the mean-squared prediction error. The model with a single explanatory variable ($x_1^*$) however offers a comparable performance. The standard Lasso rather selects the model with $x_1^*, x_2^*, x_3^*$ ($\gamma=0.06$ minimises the mean-squared prediction error). The prediction errors are, on average, smaller when using the Linear Lasso model with one or two variables than the 3-variable Lasso model. Given that one wishes to work with a 3-variable model, the best options are the Linear Lasso with $m=3$ or $m=5$.

Table \ref{figure:LS} presents the least squares estimates $c_s^t C_s^{-1}$ for each model proposed by the Linear Lasso with $m=3$. 
The standard errors of the estimates are obtained as the square root of $\sigma_s^2 C_s^{-1}$, where the estimate of $\sigma_s^2$ is the residual sum of squares, divided by $n-s$. 
Overall, both methodologies seem to agree that the first explanatory variable (police funding) has a large effect, while the other variables (all related to the population's education level) have small or moderate effects. This indicates that more police resources are allocated in cities with higher crime rates.  



\begin{table}
\begin{center}
\begin{tabular}{l|cc|cc|cc|cc|cc}
&\multicolumn{2}{|c|}{{\bf $s=5$}} & \multicolumn{2}{|c|}{{\bf $s=4$}} & \multicolumn{2}{|c|}{{\bf $s=3$}} & \multicolumn{2}{|c|}{{\bf $s=2$}} & \multicolumn{2}{|c}{{\bf $s=1$}} \\*[5pt]
 & $\hat{\beta}$ & SE & $\hat{\beta}$ & SE &  $\hat{\beta}$ & SE &  $\hat{\beta}$ & SE &  $\hat{\beta}$ & SE  \\*[5pt] \hline
$\mathbf{x}^*_1$ & \footnotesize{0.5163} & \footnotesize{0.1431} & \footnotesize{0.5326} & \footnotesize{0.1360}  &   \footnotesize{0.4893}  &  \footnotesize{0.1275} &   \footnotesize{0.4792} &  \footnotesize{0.1260}  & \footnotesize{0.5332} & \footnotesize{0.1209} \\*[5pt] 
$\mathbf{x}^*_2$ & \footnotesize{0.2064} & \footnotesize{0.2194} & \footnotesize{0.1449} & \footnotesize{0.1567}  & -- & --  & -- &  -- & -- & -- \\*[5pt]
$\mathbf{x}^*_3$ & \footnotesize{0.1123} & \footnotesize{0.2037} & \footnotesize{0.1287} & \footnotesize{0.1978}  & \footnotesize{0.2395} & \footnotesize{0.1572}  & \footnotesize{0.1732}  &  \footnotesize{0.1260} & -- & -- \\*[5pt]
$\mathbf{x}^*_4$ & \footnotesize{-0.0190} & \footnotesize{0.2199} & \footnotesize{-0.0797}  & \footnotesize{0.1593}  & \footnotesize{-0.1107} & \footnotesize{0.1555}  & -- &  -- & -- & -- \\*[5pt]
$\mathbf{x}^*_5$ & \footnotesize{-0.0965} & \footnotesize{0.2386} & -- & -- & -- & -- &  --  &  -- & -- & -- \\*[5pt] 
\end{tabular}
\caption{Least squares estimates and their standard errors for each subset $J_s$ selected by the Linear Lasso with $m=3$ and $m=5$. } \label{figure:LS}
\end{center}
\end{table}

\section{Example: Mathematics grades} \label{section:grades}

As a second example, we study student performance in secondary institutions using the dataset in \cite{cortez2008}. This dataset studies the final mathematics grades of $n=395$ students using 32 potential explanatory variables that include past student grades as well as other factors including demographic, social and school related features (age, family status, absences, etc). Nominal variables such as the field of the mother's job were converted into binary variables; the total number of variables is thus $r=41$. 

We study three different scenarios: in Scenario A, the first- and second- period grades are available ($r=41$); in Scenario B, the first-period grades are available, but the second-period grades are not ($r=40$); in Scenario C, the period grades are not available 
($r=39$). 

To find the optimal model in terms of prediction, we run a repeated 5-fold cross-validation algorithm similar to that described in the previous section. The 395 observations are thus randomly divided into 5 groups of size 79; once each of these 5 groups have acted as the test set (the other 4 groups being used to fit the model), new groups are formed and the approach is repeated a total of 50 times. This repeated cross-validation approach is first used along with the Linear Lasso (with $m=r$ and then $m=0$), each generating a nested sequence of models ranging from $r=1$ to $r=41$. The approach is then repeated with the standard Lasso using a $\gamma$-vector of length 400 in order to find the best possible model.

The cross-validation method described above generates 50 mean-squared prediction errors for each model tested. We then compute the average and standard deviation or each vector. 
Table \ref{table:grades} reports details about the best models obtained from the Linear and standard Lasso, respectively (i.e.~the models that minimize the mean-squared prediction error for each method and each Scenario A, B, and C). 

When past grades are available (first and/or second period), the models obtained show a good prediction potential. When past grades are excluded from the model, it becomes quite difficult to predict final grades, which is in line with the conclusions of \cite{cortez2008}. There is nonetheless a few variables that are kept in the model, such as the number of past failures.   

Prediction errors are similar under the Linear and standard Lasso approaches. The standard Lasso however systematically keeps a large number of explanatory variables in the model,  paradoxically offering a fit that is no better than that of the Linear Lasso in terms of prediction. The simple Linear Lasso, in contrast, offers parsimonious fits, which agrees with the claim in \cite{cortez2008} about the high number of irrelevant variables. 

Linear Lasso keeps first- and second-period grades as explanatory variables in Scenario A, first period grade and number of failures in Scenario B, and number of failures and mother's education in Scenario C. In Scenario A, Lasso keeps past grades, age, number of failures, quality of family relationships and number of absences. In Scenario B, it keeps first period grade, age, number of failures, and number of absences, but replaces other variables by existence of a romantic relationship, reason for choosing the school, etc. In Scenario C, Lasso keeps too many variables to enumerate all of them;  we however note that the number of failures is still there and has the largest coefficient, followed by the gender.

\begin{table}
\begin{center}
\begin{tabular}{l|ccc|ccc|ccc|}
&\multicolumn{3}{|c|}{{\bf A}} & \multicolumn{3}{|c|}{{\bf B}} & \multicolumn{3}{|c|}{{\bf C}}  \\*[5pt]
 & MSE & SE & $s$ & MSE & SE & $s$  & MSE & SE & $s$   \\*[5pt] \hline
Lin. Lasso ($m=r$) & \footnotesize{0.1792} & \footnotesize{0.0008} & \footnotesize{2} & \footnotesize{0.3563}  &   \footnotesize{0.0026}  &  \footnotesize{2} &   \footnotesize{0.8708} &  \footnotesize{0.0092}  & \footnotesize{2} \\*[5pt] 
Lin. Lasso ($m=0$) & \footnotesize{0.1794} & \footnotesize{0.0043} & \footnotesize{5} & \footnotesize{0.3580}  &   \footnotesize{0.0101}  &  \footnotesize{4} &   \footnotesize{0.8760} &  \footnotesize{0.0043}  & \footnotesize{1} \\*[5pt] 
Stand. Lasso & \footnotesize{0.1785} & \footnotesize{0.0019} & \footnotesize{6} & \footnotesize{0.3514}  &   \footnotesize{0.0052}  &  \footnotesize{9} &   \footnotesize{0.8777} &  \footnotesize{0.0134}  & \footnotesize{21} \\*[5pt] 
\end{tabular}
\caption{CV mean squared errors and their standard deviation for the best model of the Linear and standard Lasso in each of Scenarios A, B, and C. The number of parameters in each model is also specified; for the standard Lasso, $s$ is the number of regression coefficients greater of equal to 0.01.} \label{table:grades}
\end{center}
\end{table}

\section{Discussion} \label{section:discussion}
The Linear Lasso uses reexpressed explanatory vectors that have been sign adjusted so that  each is  positively  correlated with the interest variable; this is entirely notational but means that the explanatory vectors as so recorded will all point into the positive half space 
${\mathcal L}^{+}\bf y$, ``above"  the plane ${\mathcal L}^{\perp}\bf y$. This allows certain characteristics to be more easily described in geometric terms; and also argues that the Lasso objective itself should be 
recast as 
the scalar change    $y$ rather than as the vector change in $y\bf y$ itself.

The modified objective means that the maximum likelihood value is now on the line ${\mathcal L}\bf y$ and all the explanatory vectors intersect 
that
line at the origin. A penalty function   then becomes the  one-sided moment generating function 
$\gamma \Sigma \beta_i$
with the ``other-side" being handled by the usual positive regression coefficient requirement. As a result  computation is strictly on the line ${\mathcal L}\bf y$;
then as $\gamma$ is increased the ${\bf x}_i$ vectors are shifted in the $-{\mathcal L}\bf y$ direction, and   dropped from the lower end as determined.


When a particular  ${\bf x}_i$ is dropped 
in computation 
there is a minimum
reduction in the variance of the accessible $y$ information.
But  when a group of ${\bf x}_i$ is dropped there is no assurance that the composite change is minimum.
This is the same for the usual Lasso as it is here for the present Linear Lasso, and would be as expected from the exponential ordering in the 
possible selection of subsets.

The Linear Lasso handles cases with $r >> n$ in a straight-forward manner, by simultaneously dropping several variables featuring the smallest $c_i$'s.
It also works
generally 
with singular matrices $C$, 
due to the simplicity of the minimizing procedure.
Empirical evidence  from the real dataset examples of \S\ref{section:crime} and \S\ref{section:grades}  show that the performance of Linear Lasso is in accordance with the theoretical results developed in earlier sections; in these examples, the Linear Lasso finds models that are comparable to those found by the usual Lasso in terms of prediction accuracy, yet it consistently proposes more parsimonious models. The main advantage of Linear Lasso stems from
its
simplicity and ease of application, translating into a computational problem that is basically independent of the dimension 
once correlations are obtained.  

The theory and objective function of the Linear Lasso have been proposed in a context of multiple linear regression. Resolution algorithms for the usual Lasso offer a numerical solution for a wide range of regression models. It will be interesting to find out how the geometric arguments of the Linear Lasso can be adapted to suit other 
Lasso type contexts of interest.

\section{Appendix}
\subsection{Normal location: a tilt is a shift}
For a standard Normal $c\exp\{-z^2/2\}$ on the real line let $ \exp\{\gamma z\}$ be a factor that gives an exponential tilt or boost to the right:
$$
c\exp\{-z^2/2\}\exp\{\gamma z\}=c\exp\{-(z-\gamma)^2/2\}.
$$
We  thus see that a $\gamma$  tilt to the right can  be viewed as a $\gamma$ shift of the distribution to the right. 
Now  consider a standard Normal $c\exp\{-\sum z_i^2/2\}$ on a vector space space
coupled with a  $\gamma$ tilt in some direction $\bf x$:
$$
c\exp\{-\Sigma z_i^2/2\} \exp\{\gamma \Sigma {z_ix_i}\}=\exp\{-\Sigma (z_i -\gamma x_i )^2/2\}.
$$
Then  similarly  we see that a $\gamma$ tilt in the direction $\bf x$ can be viewed as a $\gamma$ shift  in the direction $\bf x$.

If only $s$ of some explanatory variables are being considered, say those with subscripts in the set $J_s=\{j_1,\cdots,j_s\}$, we can
use the corresponding correlation arrays as say $c_{s}, C_{s}$, and then have distributional  results analogous to the two preceding 
displays but in the appropriate subspace.


\bibliographystyle{chicago} 
\bibliography{paper-ref} 

\end{document}